\documentclass[a4paper,9pt]{article}

\usepackage{graphicx}
\usepackage{amsmath}

\usepackage{amssymb}

\newcommand{\bea}{\begin{eqnarray}}
\newcommand{\eea}{\end{eqnarray}}

\newcommand{\pdsr}{pseudosingular}
\newcommand{\pds}{pseudosingularity}
\newcommand{\pdss}{pseudosingularities}

\begin{document}

\newpage
\setcounter{page}{1}
\begin{center}
\textbf{\large{Pseudosingularity in the eigenvalue integral equations}}
\\[6ex]

{Jiao-Kai Chen 
\footnote{chenjk@sxnu.edu.cn, chenjkphy@yahoo.com}
}\\[3ex]

{\it
{\small School of Physics and Information Science, Shanxi Normal University,\\ Linfen 041004, China}\\
}
\end{center}

\begin{abstract}
The reliability is of the most importance when employing a numerical method to solve the eigenvalue integral equations. In this paper, we present one type of particular singularities (pseudosingularities) existing in eigenvalue integral equations which will impair even destroy the reliability of the numerical solutions in an implicit way. Two odd phenomena emerging in the numerical eigenvalues and the corresponding eigenfunctions are reviewed. And the relations between the pseudosingularities, the odd phenomena and the reliability of the obtained numerical results are discussed.
\end{abstract}

\vspace*{2 ex}

\begin{flushleft}
\end{flushleft}


\section{Introduction.}
The momentum-space bound-state and scattering equations are of great importance. Except for the simplest cases which have analytic solutions, these integral equations should be solved by numerical methods. When employing a numerical method to solve one integral equation, reliability, the analysis of the error in the computed results, is undoubtedly the foremost consideration and the most important issue. Reliability is addressed by a-priori and a-posteriori error estimates and error bounds. Error estimates and error bounds are related obversely to the reliability of the yielded solutions, whereas odd phenomena are related reversely to the reliability, which always appear in unexpected ways. The odd phenomena\footnote{In this paper, the considered odd phenomena arise not from the intrinsic properties of the discussed system but from the inappropriate treatment of the integral equation.} \cite{oddph} usually connect with the unreliability of the computed results and serve as an indicator of the unreliability, such as the Runge's phenomenon \cite{runge} which shows that going to higher degrees does not always improve accuracy when using polynomial interpolation to approximate one given function, the Gibbs phenomenon \cite{gibbs} which reflects the difficulty inherent in approximating a discontinuous function by a finite series of continuous sine and cosine functions, and the furcation phenomenon\footnote{When the eigenvalue integral equation which is pseudosingular along the diagonal line is solved directly by employing the quadrature method with repeated unequal weights, the obtained eigenfunctions will be in a zigzag pattern when the points are joined, while divide into branches when the point are not joined, see Fig. \ref{fig:fur1}. To distinguish it from the bifucation in the nonlinear problem, we call this peculiar phenomenon \emph{furcation}. More details are referred to Refs. \cite{Chen:2012sv,chen:2013fbs}.} \cite{Chen:2012sv,chen:2013fbs} which indicates the possible bad behaviors of the eigenvalue integral equation and the unreliability of the calculated results.

The common singularities in the integral equation are analytical, explicit and well-known to us while some peculiar singularities (\pdss{}) are numerical, implicit and concealed. The \pds{} will impair the accuracy of the numerical solutions, sometimes even destroys the reliability of the results. In most cases, this kind of unreliability can not be aware of due to its obscurity until odd phenomena are observed and identified which result from the \pds{} and indicate the unreliability of the computed results.

The paper is organized as follows. In Sec. \ref{sec:pseudo}, the \pdss{} are presented and the relation between the \pdss{} and the reliability of the numerical solutions of the eigenvalue integral equations are discussed. The relations between the \pdss{} and two odd phenomena emerging in the numerical eigenfunctions and in the eigenvalues are discussed in Sec. \ref{sec:oddphen}. The summary is in Sec. \ref{sec:conc}.

\section{Pseudosingularity and unreliability.}\label{sec:pseudo}

\subsection{Pseudosingularity}
In this paper, we concentrate on the momentum-space bound-state equation, but some conclusions will be general. For simplicity, we write the eigenvalue integral equation as
\bea\label{inteo-orig}
E\phi(p)=\int K(p,p')\phi(p')p'^2dp',
\eea
where $E$ is an eigenvalue and $\phi(p)$ is the corresponding eigenfunction, $p'^2$ is the weight for the partial wave expansion of the three-dimensional integral equation and it becomes unity when the problem is one-dimensional. In Eq. (\ref{inteo-orig}), there are different kinds of integrals. $\int$ is used to refer to any definition of the integral such that
\bea\label{hardm}
\int_a^b\frac{f(x)}{|x-y|^s}dx, \,\,\,s>0,
\eea
exists, such as the usual Riemann integral ($s<1$), the Cauchy-principal value integral ($s=1$), and the Hadamard finite-part integral ($s\ge 2$) \cite{Kutt:1975fp}. The usual singularities in the kernels are explicit, such as the logarithmic singularity in the Coulomb potential, the Cauchy-type singularity in the contact and crack problems in solid mechanics, the hypersingularity in the Cornell potential, and so on. These singular problems have been treated analytically and numerically in a large amount of literature.

Comparing with the singular Eq. (\ref{inteo-orig}), we focus on the eigenvalue integral equation of the form
\bea\label{inteo}
E\phi(p)=\int K(p,p',\beta)\phi(p')p'^2dp',
\eea
where $\beta$ represents all the other parameters. Corresponding the singular integral (\ref{hardm}), the emphasized integral maybe takes the form without loss of generality
\bea\label{hardmp}
\int_a^b\frac{f(x)}{\left[(x-y)^2+\beta^2\right]^{s'}}dx, \,\,\,s'>0.
\eea
When $\beta$ is equal to the critical value $\beta_c$ ($\beta_c=0$ for Eq. (\ref{hardmp})), Eq. (\ref{inteo}) becomes the singular Eq. (\ref{inteo-orig}). When $\beta\ne\beta_c$, Eq. (\ref{inteo}) is free of singularity strictly, but the numerical solutions of Eq. (\ref{inteo}) behave badly as $\beta$ is in the vicinity of $\beta_c$ \cite{Chen:2012sv,chen:2013fbs,Chen:2013hna} because the kernel $K(p,p',\beta)$ is not well treated in numerical integration which leads to large error when the numerical method is employed to solve Eq. (\ref{inteo}) as usual. Eq. (\ref{inteo}) with the screened Coulomb potential and with the screened Cornell potential are two concrete instances of being \pdsr{}. Analytically, the kernel $ K(p,p',\beta)$ in Eq. (\ref{inteo}) is singularity-free. Numerically, the kernel in Eq. (\ref{inteo}) is singular and should be handled just like the singular kernel in Eq. (\ref{inteo-orig}) to obtain accurate results. We use the term {\it \pds{}}\footnote{Just as the common singularities have different orders, different \pdss{} also have different orders \cite{chenprep}.} to refer to this particular type of singularities in Eq. (\ref{inteo}) which do not exist analytically but are indeed present numerically resulting in unreliable results. The \pds{} in kernel depends not only on $\beta$ but also on the numerical method.

Unlike the singularities in Eq. (\ref{inteo-orig}) which are explicit and obvious, the \pdss{} in Eq. (\ref{inteo}) are implicit and prone to being neglected. Due to the implicity of the \pds{}, we often fall into the illusion that the yielded numerical results are reliable and with expected accuracy. The illusion is reinforced when we find the numerical solutions are stable as we adjust the numerical and physical parameters. The odd phenomena emerging in the calculated results remind us that the numerical results are questionable.

When the \pdss{} are eliminated or weakened by the similar approach to deal with the corresponding singular Eq. (\ref{inteo-orig}), the \pdsr{} Eq. (\ref{inteo}) becomes
\bea\label{inteo-r}
E'\phi'(p)=\int K'(p,p',\beta)\phi'(p')p'^2dp',
\eea
where
\begin{align}\label{inteo-def}
K(p,p',\beta)&=K'(p,p',\beta)+\Delta K(p,p',\beta),\nonumber\\
 \phi(p)&=\phi'(p)+\Delta\phi(p),\nonumber\\
  E&=E'+\Delta E.
\end{align}
Analytically, $\Delta K(p,p',\beta)$ is equal to zero, and Eqs. (\ref{inteo}) and (\ref{inteo-r}) are identical with the identical solutions. Numerically, $\Delta K(p,p',\beta)\ne 0$ if Eq. (\ref{inteo}) is \pdsr{} because the \pdsr{} kernel $K(p,p',\beta)$ is not well approximated in numerical integration, and $\Delta K(p,p',\beta)$ increases as $\beta$ approaches the critical value $\beta_c$. In principle, the \pdsr{} integral can be obtained numerically with enough accuracy if step size is small enough because $\beta\ne \beta_c$ (see Eqs. (\ref{errm4}) and (\ref{errest}) below); in practice, very small step size will lead to too huge matrix when a quadrature method is implemented to solve the eigenvalue integral equation. Generally speaking, if Eq. (\ref{inteo}) is \pdsr{}, Eq. (\ref{inteo-r}) is different numerically from Eq. (\ref{inteo}) and the obtained numerical results will be also different; i.e., $\Delta E\ne 0$, $\Delta\phi(p)\ne 0$.

Substituting Eq. (\ref{inteo-def}) into Eq. (\ref{inteo}), then using Eq. (\ref{inteo-r}), we can obtain by simple calculations\footnote{The integral in Eq. (\ref{inteo-resu}) should be calculated not by analytical method but by the numerical method which is employed to solve the \pdsr{} Eq. (\ref{inteo}).} 
\begin{align}\label{inteo-resu}
\Delta E&\approx \int p^2dp\int {p'^2}dp' \phi'(p)\Delta K(p,p',\beta)\phi'(p')\nonumber\\
&\approx \int p^2dp\int {p'^2}dp' \phi(p')\Delta K(p,p',\beta)\phi(p),\nonumber\\
\phi(p)&=\frac{E'\phi'(p)+\int  K'(p,p',\beta)\Delta\phi(p'){p'^2}dp'}{E-\Delta K},
\end{align}
where $\phi(p)$ and $\phi'(p)$ are normalized. $\Delta K$ in the denominator in the last line is defined as an integral operator
\bea
\Delta K(\phi)=\int \Delta K(p,p',\beta)\phi(p')p'^2dp'.
\eea
From Eqs. (\ref{inteo-def}) and (\ref{inteo-resu}), we can see that the differences, $\Delta E$ and $\Delta\phi(p)$, arise from the \pds{} in the integral equation. Eq. (\ref{inteo-resu}) is consistent with the results in Ref. \cite{chen:2013fbs}, which are obtained when the Nystr\"{o}m method is applied to solve the integral Eqs. (\ref{inteo}) and (\ref{inteo-r})
\begin{align}\label{eigenold}
E_{nl}&\approx E'_{nl}+M_{nn},\nonumber\\
\phi_{nln}&=1,\quad \phi_{nli}\approx\frac{E'_{nl}}{E_{nl}-M_{ii}}\phi'_{nli},
\end{align}
where $M_{ij}$ is $\Delta K(p_i,p'_j,\beta)$ multiplied by the weights.

We can see from Eqs. (\ref{inteo-def}), (\ref{inteo-resu}) and (\ref{eigenold}) that $\Delta K$ and $M_{nn}$ are related directly with the \pdss{} in the integral equation and they result in errors in the eigenvalues and the corresponding eigenfunctions. $M_{nn}$ can be used as crude error estimate and a sign whether the original Eq. (\ref{inteo}) can be solved straightforwardly by numerical method. If $M_{nn}$ are small, Eq. (\ref{inteo}) can be solved directly, while if $M_{nn}$ are large, Eq. (\ref{inteo}) should be handled to weaken or remove the \pdss{} in it by the similar approach to cope with Eq. (\ref{inteo-orig}).

\subsection{Simpson's rule.}
We now illustrate the \pds{} with the Simpson's rule. Three points interpolation formula gives the Simpson's rule. The interpolating polynomial of degree two is given explicitly by the Lagrange interpolation formula
\bea\label{interpp}
\tilde{f}(x,\beta)=\sum_{i=0}^{2}L_i(x)f(x_i,\beta),\quad L_i(x)=\prod_{\substack{k=0\\k\ne i}}^{2}\frac{x-x_k}{x_i-x_k},
\eea
where $x_1=x_0+h$, $x_2=x_0+2h$, $h$ is step size. The error formula for the Lagrange polynomial reads \cite{kress}
\begin{align}\label{erroint}
R(x)&=f(x,\beta)-\tilde{f}(x,\beta)\nonumber\\
&=\frac{1}{3!}f^{(3)}(\xi,\beta)(x-x_0)(x-x_1)(x-x_2),
\end{align}
where $\xi$ is in $[x_0,x_2]$. Let $M_3(\beta)=\underset{x_0\le x\le x_2 }{\max}|f^{(3)}(x,\beta)|$, then there is
\begin{align}\label{errm3}
|R(x)|&\le \frac{h^3}{9\sqrt{3}}M_3(\beta).
\end{align}
Using Eq. (\ref{erroint}), the error for the integral reads
\begin{align}\label{errosimps}
R[f]&=\int^{x_2}_{x_0}\left[f(x,\beta)-\tilde{f}(x,\beta)\right]dx\nonumber\\
      &=-\frac{h^5}{90}f^{(4)}(\zeta,\beta), \quad \zeta\in[x_0,x_2].
\end{align}
Let $M_4(\beta)=\underset{x_0\le x\le x_2 }{\max}|f^{(4)}(x,\beta)|$, then there is
\begin{align}\label{errm4}
|R[f]|&\le \frac{h^5}{90}M_4(\beta).
\end{align}

\begin{figure*}[!ht]
\centering
\includegraphics[scale=0.29]{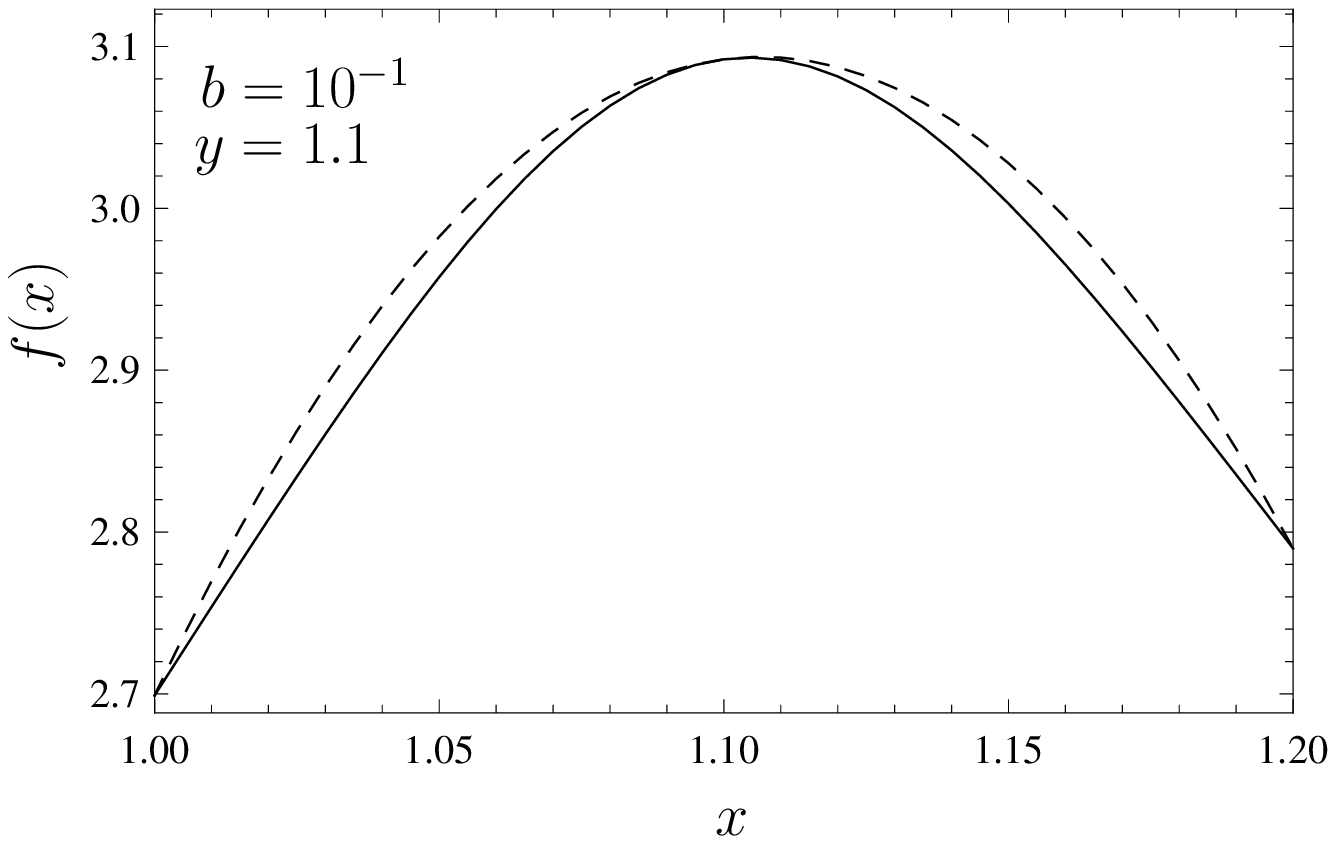}
\includegraphics[scale=0.29]{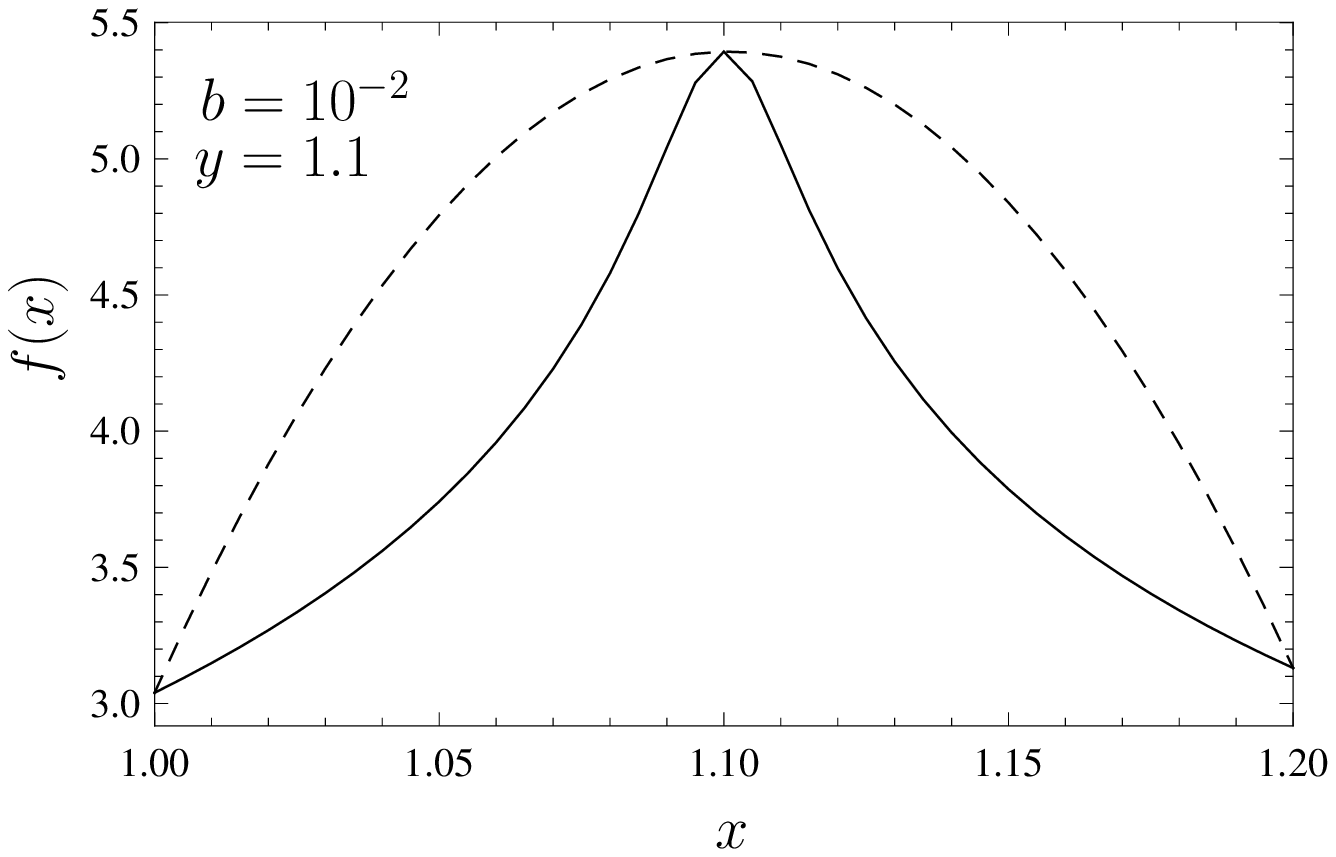}
\includegraphics[scale=0.29]{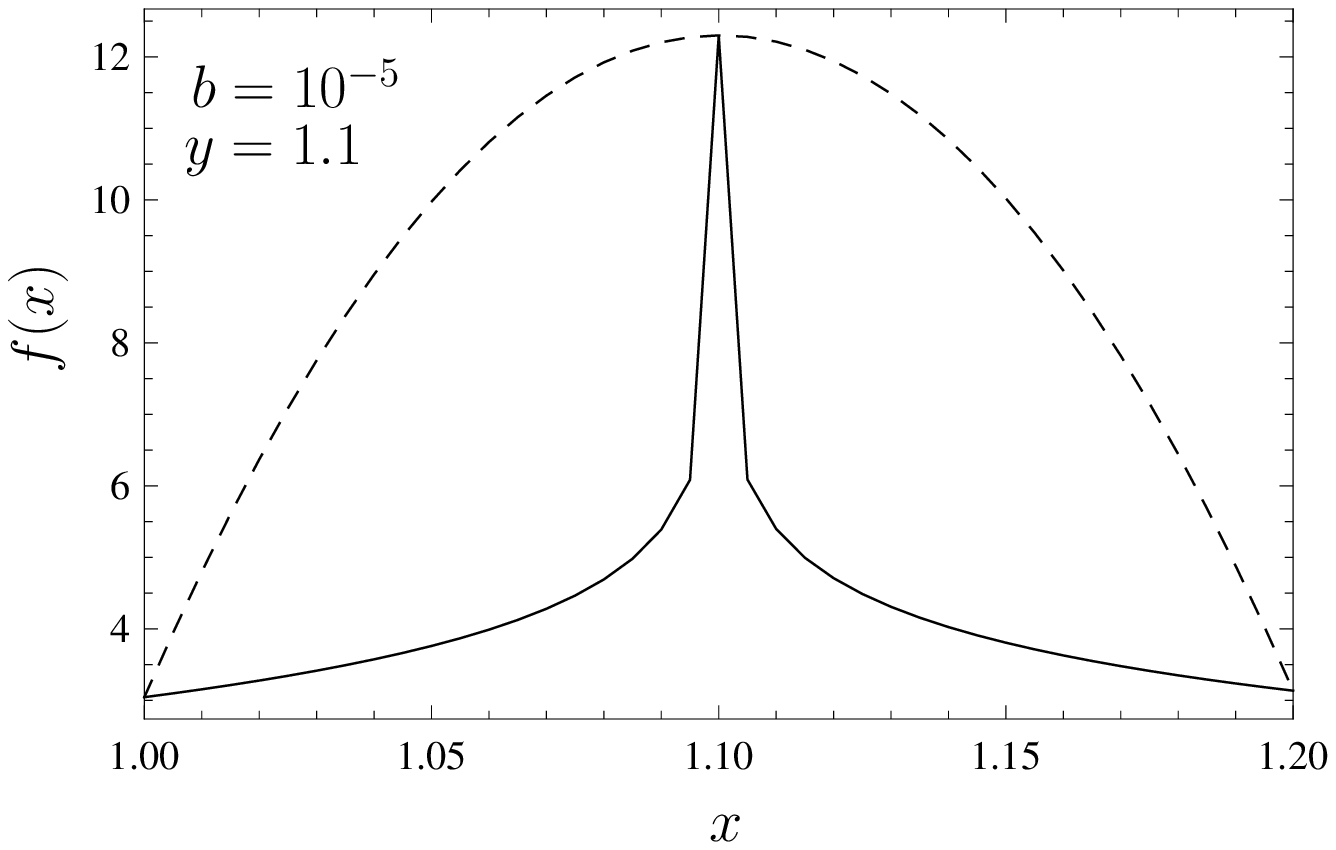}
\includegraphics[scale=0.29]{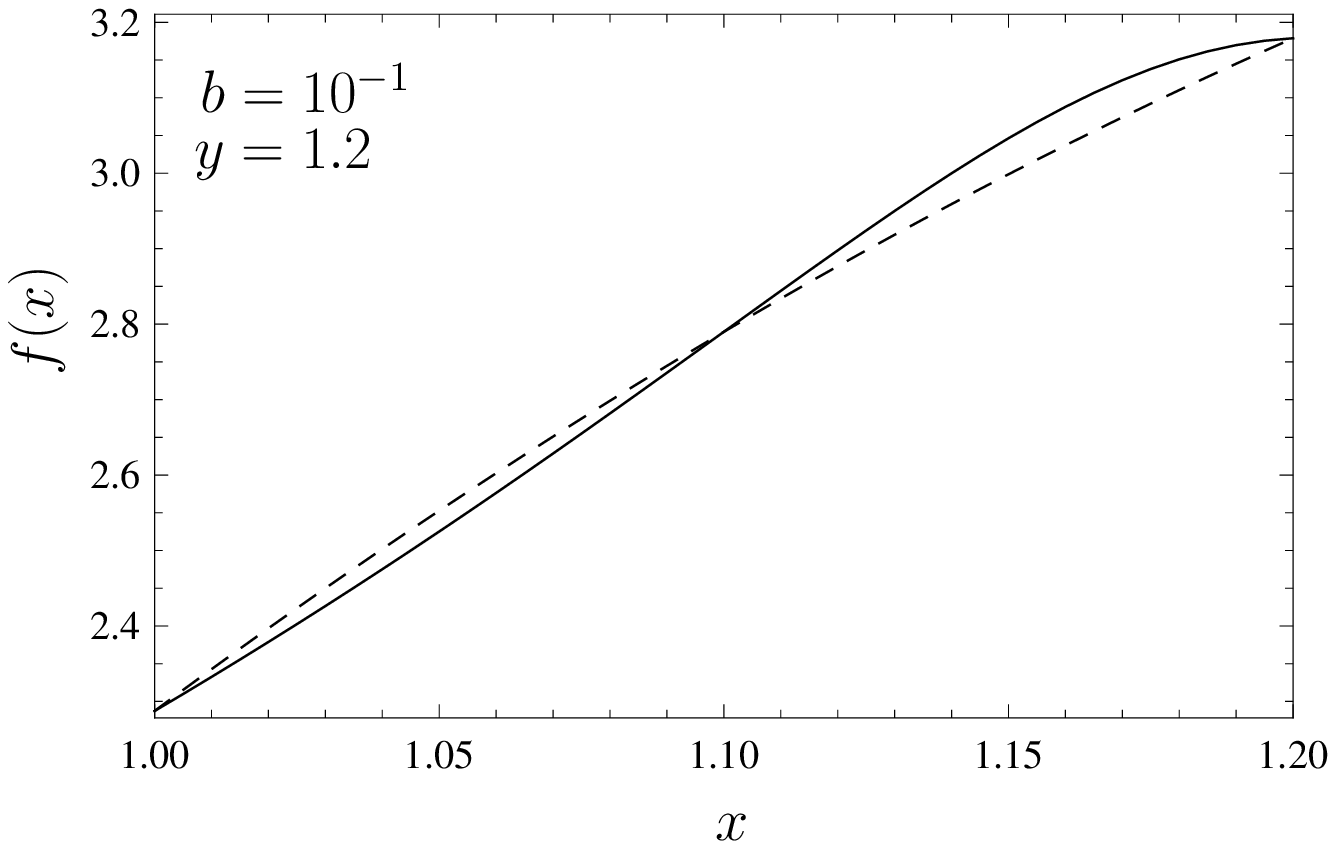}
\includegraphics[scale=0.29]{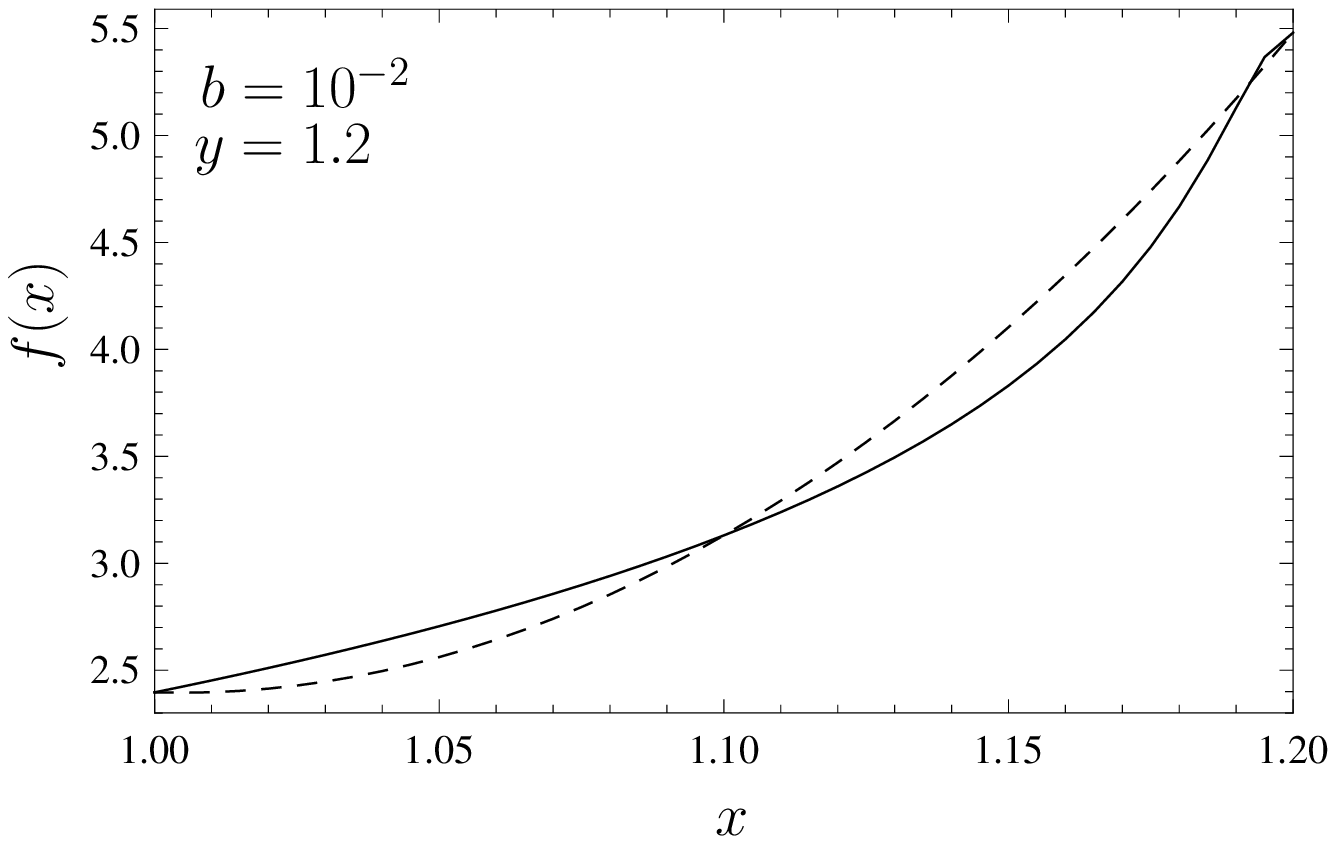}
\includegraphics[scale=0.29]{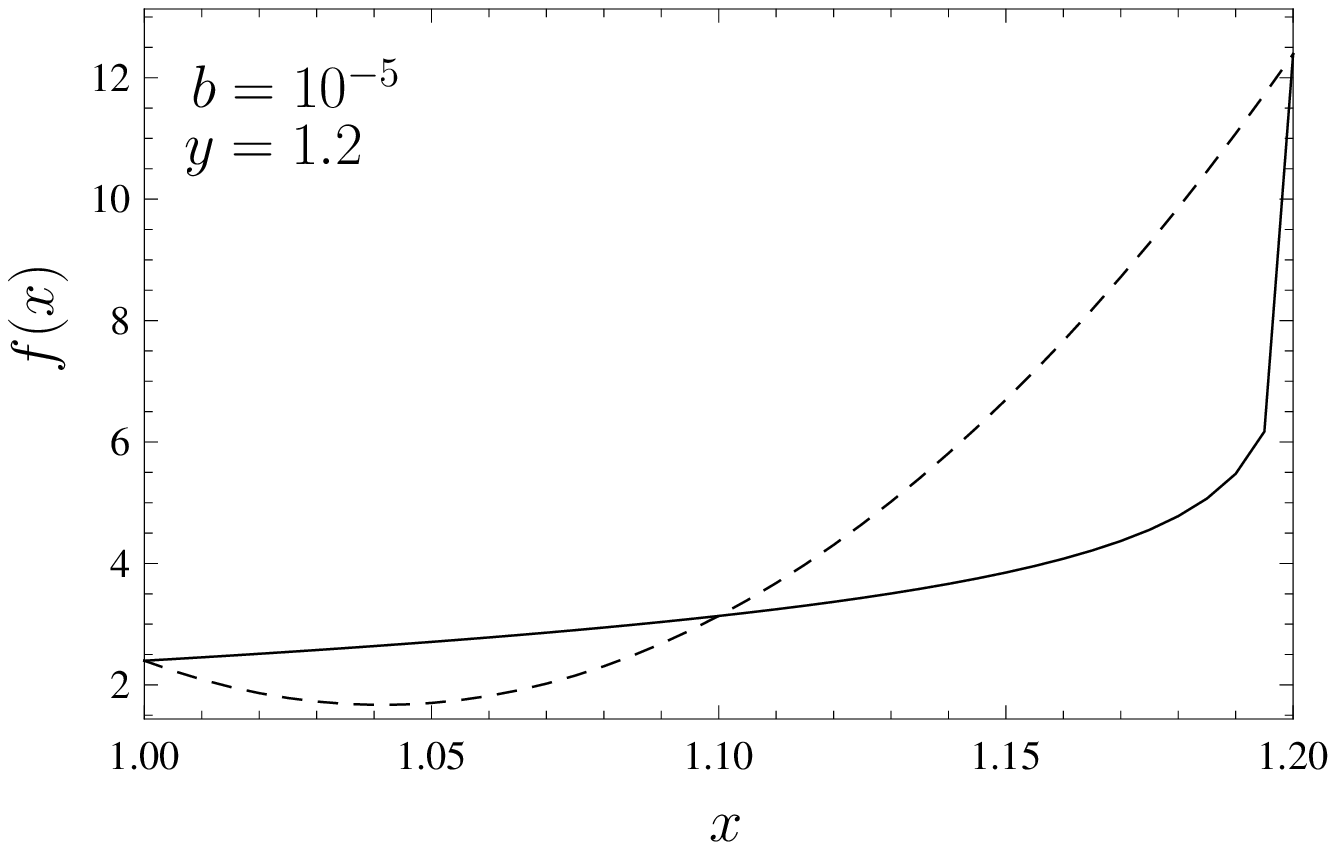}
\caption{The approximation of the function $f(x)$ (\ref{ffx}) (the solid line) by the three-point interpolating polynomial (\ref{interpp}) (the dashed line). In the upper panel, $y=1.1$, and in the lower panel, $y=1.2$.}\label{fig:simpint}
\end{figure*}

\begin{figure*}[!ht]
\centering
\includegraphics[scale=0.44]{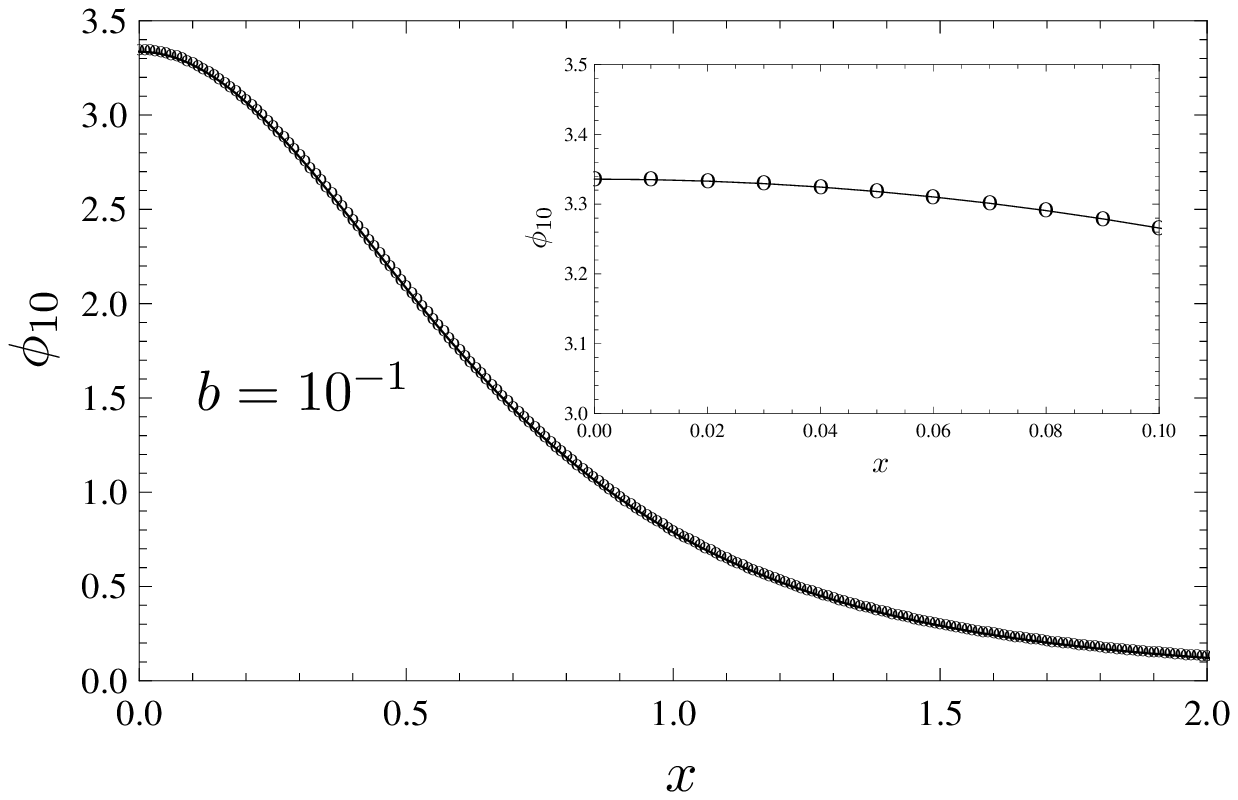}
\includegraphics[scale=0.44]{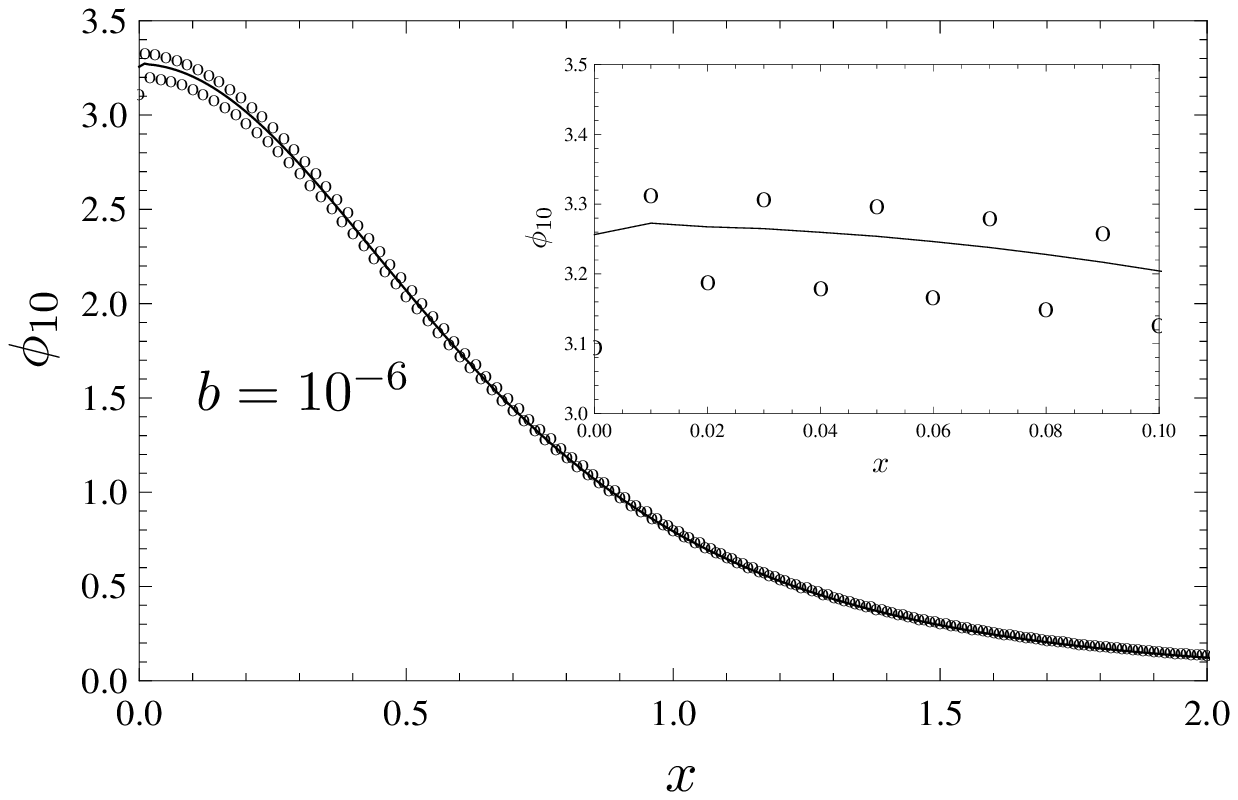}
\includegraphics[scale=0.44]{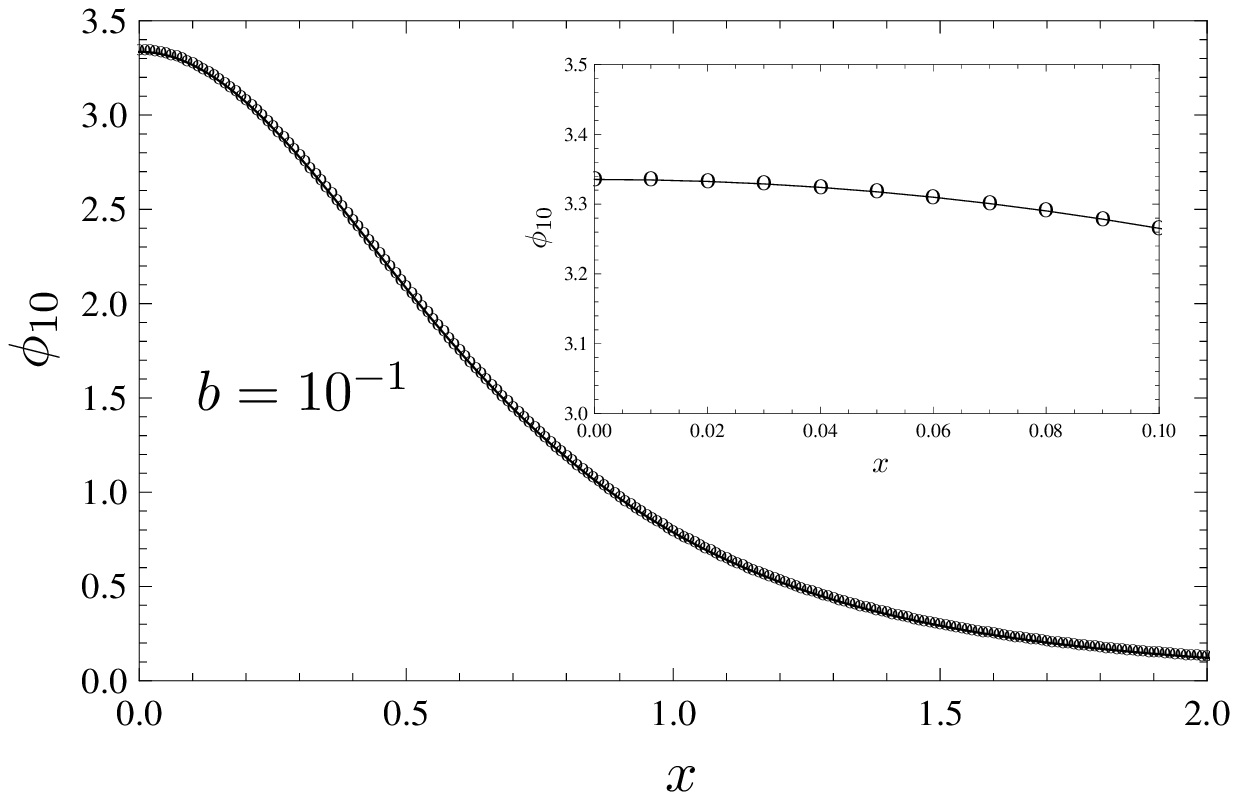}
\includegraphics[scale=0.44]{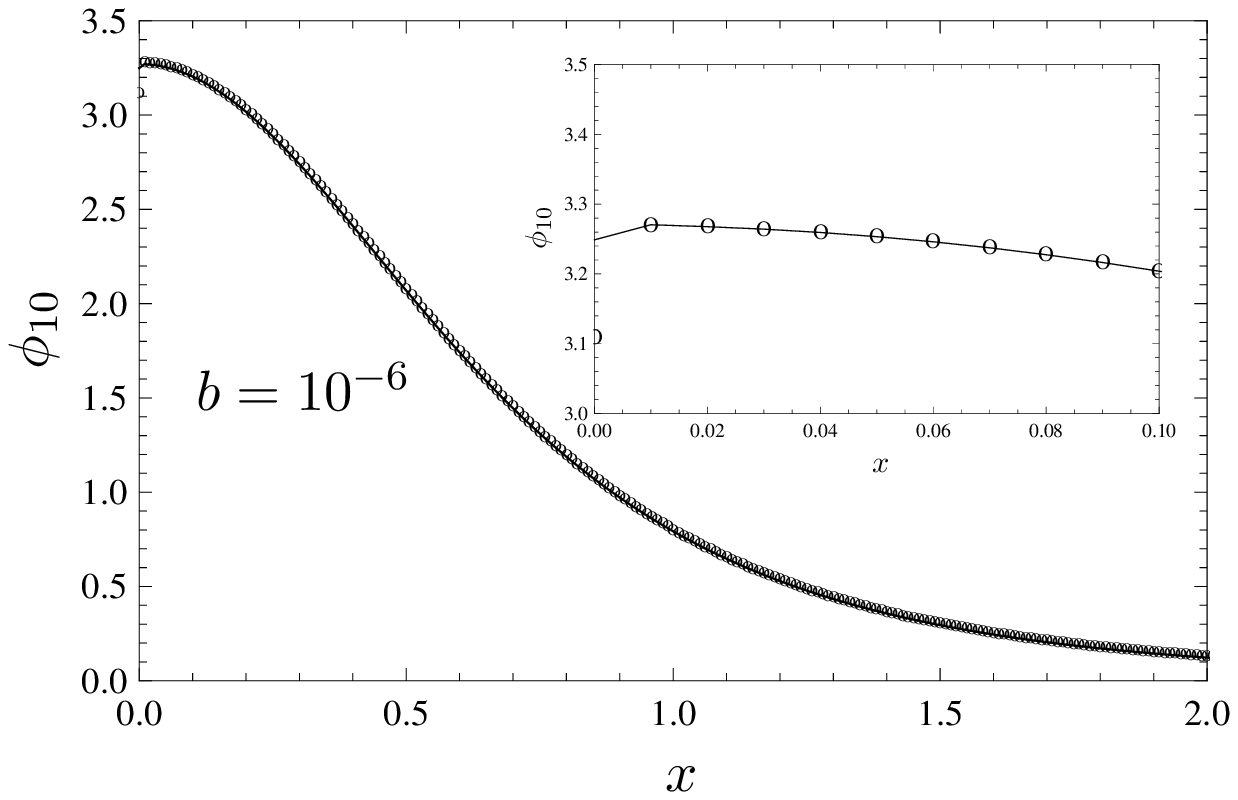}
\caption{The dependence of the furcation phenomenon on the parameter $b$ and the quadrature rules. The normalized ground state wave functions $\phi_{10}$ for the unsubtracted equation (\ref{schyuko}) (the circles) and for the subtracted equation (\ref{schyuksub}) (the solid line) are obtained numerically by applying the extended Simpson's rule (Upper panel) and by applying the extended trapezoidal rule (Lower panel). $x=p/(\mu\alpha)$, $b=\beta/(\mu\alpha)$ \cite{Chen:2012sv}.}\label{fig:fur1}
\end{figure*}

The error $R[f]$ can be divided into two parts, the part from the \pdss{} and the remains. If $f(x,\beta)$ is singular as $\beta=\beta_c$ at point $x_f$, $f(x_f,\beta)$ will be large as $\beta$ is in the vicinity of $\beta_c$, then there is the rough estimate
\bea\label{rougint}
f(x_f,\beta)\sim\ln(\Delta\beta)\;\; {\rm or}\;\; \frac{1}{\Delta\beta^{s}},\quad s>0,
\eea
where $\Delta\beta=\beta-\beta_c$. Applying Eq. (\ref{rougint}) to Eqs. (\ref{errm3}) and (\ref{errm4}) yields the asymptotic formulas at point $x_f$ or in its vicinity
\begin{align}\label{errest}
|R(x)| &\sim \frac{h^3}{\Delta\beta^{3+i}} \;\; {\rm or}\;\;  \frac{h^3}{\Delta\beta^{s+3+i}},\nonumber\\
|R[f]|&\sim \frac{h^5}{\Delta\beta^{4+i}}\;\; {\rm or}\;\;  \frac{h^5}{\Delta\beta^{s+4+i}}.
\end{align}
$i$ is related with the concrete form of the function $f(x,\beta)$. Eq. (\ref{errest}) shows that the error from the \pds{} is related to $\Delta\beta$ and step size. As $\Delta\beta$ is in the same order of or smaller than step size, the contribution from the \pds{} will dominate the error. When $\Delta\beta$ is much larger than step size, the contribution from the \pdss{} becomes small while the common error plays a dominant role. Eq. (\ref{errest}) is consistent with the numerical results, see Figs. \ref{fig:simpint}, \ref{fig:fur1}, and Table \ref{tab:schsing}. Eq. (\ref{errest}) implies that the accuracy of the numerical solutions of the \pdsr{} eigenvalue equations can be improved by reducing step size. For expansion methods, using more quadrature points \cite{atkinson} is appropriate; while for the Nystr\"{o}m methods, using very small step will be inappropriate in practice because of the resulted huge matrixes.

Applying the Nystr\"{o}m method with the extended Simpson's rule directly to Eq. (\ref{inteo}) gives
\bea\label{comsimp}
E\phi_i=\sum_j K_{ij}\phi_j p_j^2w_j+R_i, \quad R_i=R_{is}+R_{io},
\eea
where $\phi_i=\phi(p_i)$, $K_{ij}=K(p_i,p_j,\beta)$, $w_j$ are the quadrature weights, and $R_i$ is the remainder term for the $i$-th row. $R_{is}$ is the error arising from the \pdss{}, and $R_{io}$ is the left part of $R_i$. Using Eq. (\ref{errosimps}), $R_3$ can be written as
\begin{align}\label{inerr}
R_i=\sum_j-\frac{h^5}{90}G^{(4)}(\zeta_j,\beta),\quad
 G^{(4)}=\frac{\partial^{4}}{\partial{}p'^4}\big[K(p_i,p',\beta)\phi(p')p'^2\big],
\end{align}
where $\zeta_j\in[p_{j},p_{j+2}]$. From Eqs. (\ref{errest}) and (\ref{inerr}), we have that at fixed point, $\zeta_k=p_f$, $p_f\in[p_{k},p_{k+2}]$, $G^{(4)}(\zeta_k,\beta)$ will be large and behaves like ${h^5}/{\Delta\beta^{4+i}}$ or ${h^5}/{\Delta\beta^{s+4+i}}$ as $\Delta\beta$ is small. Consequently, $R_{is}$ is large and the common error $R_{io}$ becomes relatively small. Therefore, the \pdsr{} Eq. (\ref{inteo}) can not be solved directly by quadrature methods, in which the \pdss{} need to be handled. As the quadrature method case, the accuracy of the numerical solutions yielded by the expansion method is also influenced by the \pdss{}.

Taken as an example, we approximate the function $f(x)$ which is part of the integrand in the \pdsr{} Eq. (\ref{inteo}) with the screened Coulomb potential
\begin{eqnarray}
f(x)=\frac{1}{2}\ln\frac{(x+y)^2+b^2}{(x-y)^2+b^2},\label{ffx}
\end{eqnarray}
by the interpolating polynomial [Eq. (\ref{interpp})]. Without loss of generality, we set $x_0=1.0$, $x_1=1.1$, $x_2=1.2$, $y=1.1,\;1.2$. Figure \ref{fig:simpint} shows the dependence of interpolation on the screening parameter $b$. When $b$ is large, the function $f(x)$ can be well approximated by the interpolating polynomial [Eq. (\ref{interpp})], therefore, the approximation of the integral of Eq. (\ref{inteo}) by the extended Simpson's rule will behave well; while when $b$ becomes smaller and smaller, the function $f(x)$ grows more and more \pdsr{}, and then the approximation will behave worse and worse, see Fig. \ref{fig:simpint}. Therefore, as $b$ approaches the critical value $b_c=0$, Eq. (\ref{inteo}) with the screened Coulomb potential is \pdsr{} and needs to be treated just like the singular Eq. (\ref{inteo-orig}) with the Coulomb potential. This conclusion is universal and other \pdss{} in integral equations should also be treated. The existence of the \pdss{} implies that introducing parameters to control the singularities in the integral equations or to regulate the infrared divergence and avoiding the fixed point as an abscissa will fail to obtain the reliable numerical results with high accuracy.

\section{Odd phenomena}\label{sec:oddphen}
From Eqs. (\ref{inteo-resu}), (\ref{eigenold}) and (\ref{errest}), it is apparent that the \pdss{} in kernel will result in error. But if we cannot be aware of the existence of the \pdss{} in the integral equation, generally, we cannot discern the error of the numerical solutions from the \pdss{} and the error from other sources. This situation is partly due to the stability of the numerical results when fine-tuning the parameters.

The \pdss{} and the resulting errors are implicit and easy to be neglected. Fortunately, there are odd phenomena emerging in the numerical eigenfunctions and in the eigenvalues which can be used as indicators to throw light on the unreliability of the numerical results of the \pdsr{} integral equation.

\begin{table*}[!ht]
\caption{The eigenvalues of the \pds{}-free Schr\"{o}dinger equation (\ref{schyuksub}) (C) which are regarded as reliable results and of the \pdsr{} Schr\"{o}dinger equation (\ref{schyuko}) (NC) with the screened Coulomb potential are calculated by employing the Nystr\"{o}m method with the extended Simpson's rule (SR) and with the extended trapezoidal rule (TR). $b=\beta/(\mu\alpha)$ is the redefined screening parameter. The data are from Table \uppercase\expandafter{\romannumeral3} in Ref. \cite{Chen:2012sv}.}\label{tab:schsing}
\centering
\begin{tabular*}{\textwidth}{@{\extracolsep{\fill}}cc llll@{}}
\hline\hline
b &n  & TR(C)&TR(NC)          & SR(C)&SR(NC)          \\
\hline
        &1  &-0.397332         &-0.397331 &-0.397332      &-0.397331\\
$10^{-1}$&2 &-0.0488747        &-0.0488742&-0.0488746      &-0.0488742\\
        &3 &-0.00306981       &-0.00306958&-0.00306981    &-0.00306958\\
        &1 &-0.489129         &-0.491555  &-0.489129      &-0.492217  \\
$10^{-3}$&2 &-0.122730         &-0.125155  &-0.122731      &-0.125835  \\
         &3  &-0.0541873        &-0.0566106  &-0.0541916      &-0.0573194 \\
         &1  &-0.490127         &-0.513594  &-0.490127       &-0.51679   \\
$10^{-6}$&2  &-0.123726         &-0.147192  &-0.123727      &-0.150749  \\
         &3  &-0.0551799        &-0.0786442 &-0.0551842     &-0.0827475\\
\hline\hline
\end{tabular*}
\end{table*}

\subsection{Furcation}

When the Nystr\"{o}m method with the extended Simpson's rule is employed to solve the integral Eq. (\ref{inteo}) which is \pdsr{} along a line (as an example, the Schr\"{o}dinger equation with the screened Coulomb potential is listed in \ref{app:sch}), the furcation phenomenon \cite{Chen:2012sv,chen:2013fbs} will emerge in the eigenfunctions, see Fig. \ref{fig:fur1}. From Eqs. (\ref{inteo-resu}) and (\ref{eigenold}), we can see that the alternate wobble of the weights for the extended Simpson's rule, $1/3$, $4/3$, $2/3$, $\cdots$, leads to the alternate wobble of $M_{ii}$ which results in the wobble of the point of the yielded eigenfunction---that is, the furcation with two branches emerges. The furcation phenomenon with three branches occurs in the numerical eigenfunctions \cite{chen:2013fbs} when employing the Nystr\"{o}m method with the extended closed Newton-Cotes formula of degree $5$. The furcation phenomenon indicates the unreliability of the numerical results and the existence of the \pds{} in the integral equation.

According to Eqs. (\ref{inteo-resu}) and (\ref{eigenold}), different numerical methods will make the calculated results behave badly in different manner. For example, when the extended Simpson's rule is employed, the eigenvalues are of low accuracy and the furcation phenomenon emerges in eigenfunctions, while when the extended trapezoidal rule is implemented, there is not furcation phenomenon but the accuracy of eigenvalues is still low, see Fig. \ref{fig:fur1} and Table \ref{tab:schsing}. How can we know whether the obtained results are good or bad? In this sense, the emergence of the furcation phenomenon in the calculated eigenfunctions which can give an evident warning is not the defect of the quadrature rules with repeated unequal weights but their merit.

\subsection{Abnormal convergence direction}

If the exact eigenvalues belonging to the kernel $K$ are monotonic and have the same sign,
\bea\label{convdir}
 E_{min}<E_1<\cdots<E_n<\cdots<E_{max},
\eea
the reliable numerical eigenvalues should satisfy the following relations \cite{Chen:2013hna,delv,baker} no matter what kind of numerical method is employed
\begin{align}\label{eigprop}
&E_{min}\le E_{min}^{N}\le E_1^{N}\le\cdots \le E_i^{N}\cdots\le E^N_{max}\le E_{max}, \nonumber\\
&E_{min}\le E_{min}^{N+1}\le E_{min}^{N},\; \cdots,\;
E_{max}^{N}\le E_{max}^{N+1}\le E_{max}.
\end{align}
The properties [Eq. (\ref{eigprop})] of the numerical eigenvalues are of great practical importance. The results in Eq. (\ref{eigprop}) are independent of the choice of the numerical methods and therefore are general. Eq. (\ref{eigprop}) implies that the convergence of each eigenvalue is monotonic in $N$ and the direction of the convergence is definite. We call the convergence direction of a numerical eigenvalue obeying Eq. (\ref{eigprop}) a {\it normal convergence direction}; otherwise, it is an {\it abnormal convergence direction}. The reliable eigenvalues should obey the normal convergence directions. While if the obtained eigenvalue converges abnormally, it is necessary to check it, no matter the problem arises from the adopted numerical method or from the bad behavior of the eigenvalue equation to be solved or from the introduced parameters. In practical applications, Eq. (\ref{eigprop}) is valid only for states with low energy, and will be inappropriate to the highly excited states. One reason is that the convergence direction becomes ambiguous for highly excited states---that is to say, it is not definite whether the numerical eigenvalues in the middle part converge to the exact values from the left side or from the right side, see Eq. (\ref{eigprop}). Another cause is that the latter part of the numerical eigenvalues will be nonphysical in many cases (see \ref{app:nonphy}).

Defining a functional $F_{K}$ in inner product notation
\bea\label{rayjud}
F_{K}(\psi)={(\psi,K\psi)}, \quad {(\psi,\psi)=1},
\eea
where $\psi$ can be obtained by analytical methods or by numerical methods. This functional is usually referred to as the Rayleigh quotient. $F_{K}(\phi^N)$ will be the corresponding eigenvalue,
\bea\label{rayjudb}
F_{K}(\phi^N)=E_{exact}+\Delta E^N,
\eea
if $\phi^N$ is the normalized eigenfunction yielded by a numerical method, where $N$ represents $N$ nodes or $N$ basis functions. $E_{exact}$ is the exact eigenvalue, and $\Delta E^N$ is the difference between the calculated eigenvalue and the exact result for this time. When more points or more basis functions are used next time, $\Delta E^N$ will change. From Eq. (\ref{eigprop}), we can see that for small eigenvalues close to the ground-state eigenvalue $E_{min}$, the numerical result $F_{K}(\phi^N)$ will converge normally if $\Delta E^N\ge 0$, while converges abnormally if $\Delta E^N< 0$.

In case of the \pdsr{} eigenvalue equation, applying the Rayleigh quotient (\ref{rayjud}) and using Eqs. (\ref{inteo}), (\ref{inteo-r}) and (\ref{inteo-def}) give
\begin{align}\label{raylfn}
F_{K}(\phi)&\approx F_{K'}(\phi')+\Delta F,
\end{align}
where
\begin{align}
\Delta F =(\phi',\Delta K\phi).
\end{align}
The above Eq. (\ref{raylfn}) is just the result in Eq. (\ref{inteo-resu}). $F_{K}(\phi)$ in Eq. (\ref{raylfn}) is the numerical eigenvalue of the \pdsr{} Eq. (\ref{inteo}) and $F_{K'}(\phi')$ is the corresponding eigenvalue of the treated Eq. (\ref{inteo-r}) which is regarded as reliable result. $\Delta F$ is related to the \pdss{} and the numerical methods. According to Eqs. (\ref{eigprop}), (\ref{rayjudb}) and (\ref{raylfn}), we can obtain that if the kernel is free of \pdss{}, $\Delta F\ge 0$ and the numerical eigenvalue converges normally, while $\Delta F< 0$ and the convergence direction will be abnormal if the integral equation is \pdsr{} which indicates the possible unreliability, see Table \ref{tab:schsing}. Table \ref{tab:schsing} shows one case of the abnormal convergence direction, the adverse convergence direction that the obtained eigenvalues converge from adverse direction when $b=10^{-3}$ and $b=10^{-6}$. Another case of the abnormal convergence direction is presented in \ref{app:osci}.

\section{Summary}\label{sec:conc}
In this paper, we have presented the \pdss{} in the eigenvalue integral equations, which will result in the unreliability of the numerical solutions. The \pdss{} are implicit and prone to being neglected. The emerging odd phenomena, the furcation phenomenon in eigenfunctions and the abnormal convergence direction of eigenvalues, give us warning. When the numerical anomalies are observed, it is then necessary to check the reliability of the obtained numerical solutions.

\section*{Acknowledgements}
This work was supported by the Natural Science Foundation of Shanxi Province of China under Grant No. 2013011008.

\appendix
\section{Schr\"{o}dinger equation with the screened Coulomb potential}\label{app:sch}

Taken as an example to show explicitly the furcation phenomenon, we solve the Schr\"{o}dinger equation with the screened Coulomb potential \cite{Chen:2012sv,chen:2013fbs}. The partial expansion of the Schr\"{o}dinger equation is written as
\bea\label{schyuko}
E_{nl}\phi_{nl}(p)=\frac{{p}^2}{2\mu}\phi_{nl}(p)
+ \frac{1}{(2\pi)^{3}}\int_0^{\infty}V^l_Y(p,p')\phi_{nl}(p')p'^2dp',
\eea
where $n$ is the principal quantum number, $l$ is the orbital angular quantum number. $V_Y^{l}(p,p')$ is the partial wave expanded screened Coulomb potential,
\bea
V_Y^{l}(p,p')=-8\pi^2\alpha\frac{Q_l(z)}{pp'}, \quad z\equiv\frac{{p^{2}+p'^{2}+\beta^{2}}}{{2p'p}}, \label{pypot}
\eea
where $Q_l(z)$ is the Legendre polynomial of the second kind,
\bea
Q_{l}(z) &=& P_{l}(z)Q_{0}(z)-w_{l-1}(z), \quad Q_{0}(z) = {1 \over 2} \ln \frac{z+1}{z-1},\nonumber\\
 w_{l-1}(z) &=& \sum^{l}_{m=1}{1 \over m} P_{l-m}(z)P_{m-1}(z).\label{QL}
\eea
If the screening factor $\beta=\beta_c$ where $\beta_c=0$, the screened Coulomb potential $V^{l}_{Y}(p,p')$ will reduce to the Coulomb potential which has the logarithmic singularity at point $p'=p$, and the singularity comes from $Q_{0}(z)$. If the screening factor $\beta>0$, the screened Coulomb potential is free of singularity analytically. The numerical results, however, have bad behaviors when the Nystr\"{o}m method with the extended Simpson's rule is employed to solve the integral equation (\ref{schyuko}).

Applying the Land\'{e} subtraction method \cite{norburysse} to cancel out the \pds{}, the \pdsr{} equation (\ref{schyuko}) becomes
\begin{align}\label{schyuksub}
E_{nl}\phi_{nl}(p)=&\,\frac{{p}^2}{2\mu}\phi_{nl}(p)
- {\alpha\over \pi p} \int^{\infty}_{0} P_{l}(z)\frac{Q_{0}(z)}{p'} \left[ p'^{2}\phi_{nl}(p')
- \frac{P_{l}(z')}{P_{l}(z)}p^{2}\phi_{nl}(p) \right] dp' \nonumber \\
& -\frac{\alpha p}{\pi} \left[ {\pi^{2}\over 2}-\pi \arctan\frac{\beta}{p}\right]P_{l}(z')\phi_{nl}(p)
+{\alpha\over \pi p}\int^{\infty}_{0}w_{l-1}(z)\phi_{nl}(p')p'dp',
\end{align}
where $z'=(2p^2+\beta^2)/(2p^2)$. This subtracted equation is free of singularity and \pds{} and can be applied to study both the Coulomb potential problem ($\beta=0$) and the screened Coulomb potential problem ($\beta>0$).

From Eqs. (\ref{inteo-def}), (\ref{schyuko}) and (\ref{schyuksub}), we can see that $V_Y^{l}(p,p')$ is \pdsr{} along the line $p=p'$, therefore, $\Delta K(p,p',\beta)$ is not equal to zero and will increase as the screening parameter $\beta$ approaches the critical value $\beta_c$ ($\beta_c=0$). From Eqs. (\ref{inteo-resu}) and (\ref{eigenold}), we can see that $M_{ii}$ varies with the screening parameter $\beta$ and the weights in the numerical method and leads to the error in the eigenvalues and the corresponding eigenfunctions. As shown in Fig. \ref{fig:fur1} and Table \ref{tab:schsing}, the eigenvalues and eigenfunctions for the \pds{}-free equation (\ref{schyuksub}) are good, while the obtained eigenvalues for the \pdsr{} equation (\ref{schyuko}) are of low accuracy and have abnormal convergence direction, and the corresponding eigenfunctions divide into two branches at small x when the extended Simpson's rule is applied, that is to say, the furcation phenomenon emerges in eigenfunctions. Both the furcation phenomenon in the eigenfunction and the abnormal convergence direction of the eigenvalues indicate the \pds{} in the integral equation and the unreliability in the numerical results.

\section{Nonphysical eigenvalues}\label{app:nonphy}

From Eqs. (\ref{convdir}), (\ref{rayjud}) and the relation \cite{Chen:2013hna}
\bea
F_K(\psi)\ge E_{min},\quad F_K(\psi)\le E_{max},
\eea
the eigenvalues calculated by the quadrature method, the expansion method or the spectral method, should be larger than the minimal eigenvalue and be smaller than the largest one. In practice, when using the numerical method to solve the eigenvalue integral equation whether it is singular or not, there exist not only physical eigenvalues which are reasonable but also nonphysical eigenvalues in the obtained numerical solutions, which are nonsense physically and are the latter part of the obtained eigenvalues. For example, the eigenvalues of the Schr\"{o}dinger equation with the Coulomb potential should be all negative, but some yielded results are positive. We have been familiar to and accustomed to this phenomenon.

In case of quadrature method, for large $n$, the strongly oscillating wave functions with many nodes congregating in the vicinity of the zero point cannot be well approximated by the numerical ones, which leads to the nonphysical eigenvalues $F(\psi^{N})$ and the corresponding eigenfunctions  $\psi^{N}$,
\bea\label{nysray}
F(\psi^{N})={\left(\psi^{N},K\psi^{N}\right)}
           ={\sum_{i,j=1}^{N}w_iw_j\psi^{N}_{i}K_{ij}\psi^{N}_j},
\eea
where $\psi_i^{N}=\psi^{N}(p_i)$, $K_{ij}=K(p_i,p_j)$, $w_i$ are the weights. Generally speaking, nonequal step method works better than equal step method in which the equal abscissa will leads to more nonphysical eigenvalues.

Now we discuss the expansion method case. Consider two set of basis functions, $\{\varphi_i\}$ and $\{\phi_i\}$ which are all orthonormal and complete. The latter is the set of the exact eigenfunctions for the integral equation. Due to the completeness of the two set of basis, there is the formula
\bea
\phi_i=\sum_{i=1}^{\infty} a_{ij}\varphi_j, \quad i=1,\cdots, \infty.
\eea
In practice, we must truncate it at a finite number $m$,
\bea
\phi_i\approx\sum_{i=1}^{m} a_{ij}\varphi_j, \quad i=1,\cdots, m.
\eea
If the former $n$ functions $\phi_i$ ($i=1,\cdots,n$) can be well approximated by $\varphi_i$, we can obtain by the Gram-Schmidt process
\bea
\{\varphi_1,\cdots,\varphi_n,\varphi_{n+1},\cdots,\varphi_m\}\approx \{\phi_1,\cdots,\phi_n,\varphi'_{n+1},\cdots,\varphi'_m\}.
\eea
If $\{\varphi'_j,j=n+1,\cdots,m\}=\{\phi_j,j=n+1,\cdots,m\}$, the two subsets, $\{\varphi_j,j=1,\cdots,m\}$ and $\{\phi_j,j=1,\cdots,m\}$, will be equal and span the same space. Obviously, the left $m-n$ functions $\phi_j$ cannot be well approximated by $\varphi_j$ and vice versa, therefore, the left $n-m$ eigenfunctions and the corresponding eigenvalues have large errors, sometimes the eigenvalues are beyond the physical limit and becomes nonphysical.

Nonphysical eigenvalues emerge not only in numerical results but also in analytical results when employing expansion method. Generally speaking, nonphysical eigenvalues have nothing to do with the \pdss{} in the integral equation and cannot be avoided unless the basis functions are the exact eigenfunctions for the integral equation to be solved. The choice of the basis function has an effect on the number of nonphysical eigenvalues and the accuracy of the physical eigenvalues. Thus, the number of nonphysical eigenvalues can be regarded as a not-so-good indicator of the unreliability of the numerical solutions.

In summary, in case of the quadrature method, the highly oscillating eigenfunctions for large $n$ cannot be well approximated by the numerical solutions at small abscissas. Therefore, it leads to the nonphysical eigenvalues. In case of the expansion method, nonphysical eigenvalues are yielded because the eigenfunctions for large $n$ cannot be well approximated by the basis functions. In general, the quadrature method will be better than the expansion method for the nonphysical eigenvalue problem.

\section{Oscillation of the convergence direction}\label{app:osci}
The adverse convergence direction of the obtained eigenvalues is shown in Table \ref{tab:schsing}. Another case of the abnormal convergence direction, the oscillating convergence direction, is shown in this section. Consider the eigenvalue integral equation \cite{delv}
\bea\label{delvanom}
\gamma x(s)=\int^{1}_{0}\left(st-\frac{1}{6}s^3t^3\right)x(t){\rm d}t.
\eea
Using the monomials $s^{i-1}$ as basis functions to approximate the eigenfunctions, the Ritz-Galerkin method applied to the solution of the integral equation yields a matrix equation, which can be solved using a generalized eigenvalue routine.

\begin{table}[!htb]
\caption{The largest eigenvalue for equation $\gamma x(s)=\int^1_0 (st-s^3t^3/6)x(t)dt$ calculated by an expansion method. The data are from Table 7.4.1 in Ref. \cite{delv}.}\label{tab:osci}
\begin{center}
\begin{tabular}{cc cc c}
\hline\hline
N       & 2        & 4         &8       & Exact\\
\hline
i=1  & 0.31344  & 0.31357   &0.31359 &0.31357\\
\hline\hline
\end{tabular}
\end{center}
\end{table}

In Eq. (\ref{delvanom}), the kernel is free of singularity or pseudosingularity. The numerical results should be yielded easily and be with hight accuracy. Out of expectation, the obtained eigenvalues have oscillating convergence direction, see Table. \ref{tab:osci}. The abnormal convergence direction tells us that we should be cautions of whether the results are reliable or not and should check it carefully.

As a check, we use the Mathematica package to compute the eigenvalues by employing the method applied by Delves because Mathematica can handle approximate real numbers with any number of digits. The results are
\begin{eqnarray}
 \frac{1}{420}\left(65+\sqrt{4449}\right),\quad \frac{1}{420}\left(65-\sqrt{4449}\right),\quad 0,\cdots.
\end{eqnarray}
The results have been checked to $N=200$. The obtained eigenvalues are bad when the data in lower precision while become good as the data in high precision. Therefore, the unreliability of the numerical results stems from the insufficient precision of the numerical calculation which results in the roundoff errors. In finite precision mathematics, the unreliability of the obtained results has been attributed to the inappropriate choice of basis \cite{delv}.



\begin{thebibliography}{99}
\bibitem{oddph}
Some odd phenomena exist even in the analytical results, such as the Runge's phenomenon and the Gibbs phenomenon, while others emerge only in the numerical results, for instance, the furcation phenomenon. In this paper, we concentrate on the odd phenomena occurring in the numerical solutions of the integral equation for eigenvalue problem with well-behaved exact solutions. The odd phenomena arise from the bad behavior of the integral equation and/or from the employed numerical method.
\end{thebibliography}
\end{document}